\documentclass{article}
\usepackage{arxiv}
\usepackage{ifluatex}

\usepackage{colortbl} 
\usepackage{mathptmx} 
\usepackage{helvet} 
\usepackage{courier} 
\usepackage{type1cm} 
\usepackage{makeidx} 
\usepackage{graphicx} 

\usepackage{multicol} 
\usepackage[bottom]{footmisc}

\usepackage{amsbsy}
\usepackage{amsfonts}
\usepackage{amsmath}
\usepackage{color}
\usepackage{adjustbox}
\usepackage{moreverb}

\usepackage{algorithm}
\usepackage[algo2e, boxed, ruled, vlined]{algorithm2e}
\usepackage{subfigure}
\usepackage{lmodern}
\usepackage{amssymb}
\usepackage{hyperref}       
\usepackage{url}            
\usepackage{booktabs}       
\usepackage{nicefrac}       
\usepackage{microtype}      

\providecommand{\bo}{\mathbf}

\providecommand{\cov}{\mathrm{\bo {COV}}}
\providecommand{\E}{\mathrm{\bo E}}

\newtheorem{result}{Result}
\newtheorem{definition}{Definition}

\numberwithin{equation}{section}
\title{Non-Gaussian
component analysis: testing the dimension of the signal subspace}

\author{Una Radojicic  \\
  Vienna University of Technology\\
  \texttt{una.radojicic@tuwien.ac.at} \\
  \And
        Klaus Nordhausen \\
  Vienna University of Technology\\
  \texttt{klaus.nordhausen@tuwien.ac.at} \\
  }

\renewcommand{\headeright}{}
\renewcommand{\undertitle}{}
\date{}

\begin{document}

\maketitle

\begin{abstract}
Dimension reduction is a common strategy in multivariate data analysis which seeks a subspace which contains all interesting features needed for the subsequent analysis. Non-Gaussian component analysis attempts for this purpose to divide the data into a non-Gaussian part, the signal, and a Gaussian part, the noise. We will show that the simultaneous use of two scatter functionals can be used  for this purpose and suggest a bootstrap test to test the dimension of the non-Gaussian subspace. Sequential application of the test can then for example be used to estimate the signal dimension.
\end{abstract}

	\section{Introduction}
	\label{Radojicic:sec:1}
	
	Modern data sets contain often many variables making visualization and many other tasks concerning the data set very difficult. Therefore, dimension reduction methods gain popularity as they try to find a subspace of the data which is smaller and contains all interesting features. Three main issues are then here, (i) how to define what makes the data interesting, (ii) how large is the interesting subspace and (iii) how to find the subspace?
	
	There are meanwhile many suggestions about how to define what is interesting and maybe the most used method is principal component analysis (PCA) \cite{Jolliffe2002} which defines as interesting subspace the one which accounts for as much of the variability in the data as possible. Another well-established approach is projection pursuit (PP) \cite{Huber:1985,jones1987projection} where usually univariate projections of the data, which maximize some criterion of non-Gaussianity specified by an projection index, are considered interesting. PCA is probably so popular as it is quite easy to compute and has many different guidelines on how to choose the dimension of the subspace of interest. PP on the other hand is, depending on the projection index used, often computationally expensive. Moreover, guidelines about how to choose the dimension of the interesting subspace are sparse. However, PP has been proven useful as a preprocessing step for, for example, clustering or outlier detection \cite{FischerBerroNordhausenRuizGazen2019}. In general, it seems that the non-Gaussian subspace of the data is nowadays considered the subspace of interest and \cite{Blanchard05} suggested a general framework for this, denoted by non-Gaussian component analysis (NGCA). It divides the data into a non-Gaussian subspace and into a Gaussian subspace. While there are meanwhile many suggestions, like in \cite{BlanchardKawanabeSugiyamaSpokoinyMuller2006,KawanabeSugiyamaBlanchardMuller2007,Theis:2011,Bean2014,SasakiNiuSugiyama2016,Virta:2016} to name a few, on how to perform NGCA there is not much research yet on how to estimate the dimensions of the two subspaces.
	
	In this paper we will introduce a bootstrap test to test the dimension of the non-Gaussian subspace using two scatter matrices. For this purpose we will in the following first introduce scatter matrices and some of their relevant properties. Then, in Section \ref{Radojicic:sec:3} we will introduce the independent component (IC) model which is closely related to the NGCA model, which we will also define then there in detail.
	The bootstrap test is then introduced in Section\,\ref{Radojicic:sec:4} and evaluated in a simulation study in Section \ref{Radojicic:sec:5}. Natural estimates of the signal dimension are found by successive conduction of the bootstrap test and two estimation strategies are discussed and evaluated in Section \ref{Radojicic:sec:6}. Proofs of  selected results are given in the Appendix.
	
	\section{Scatter Functionals}
	\label{Radojicic:sec:2}
	
	Scatter functionals are the main tools in our method and defined as follows:
	
	\begin{definition}
		Let $\bo x$ be a $p$-variate random vector with distribution function $F_{\bo x}$. Then a $p \times p$ matrix-valued functional $\bo S(F_{\bo x})=\bo S(\bo x)$ is called a scatter functional if it is symmetric, positive semi-definite and affine equivariant in the sense that
		\[
		\bo S(\bo A\bo x + \bo b) = \bo A \bo S(\bo x) \bo A^\top,
		\]
		for all full rank $p \times p$ matrices $\bo A$ and all $p$-variate vectors $\bo b$.
	\end{definition}
	
	Scatter functionals often come along with a location functional which is defined as:
	\begin{definition}
		Let $\bo x$ be a $p$-variate random vector with distribution function $F_{\bo x}$. Then a $p$-vector-valued functional $\bo T(F_{\bo x})=\bo T(\bo x)$ is called a location functional if it is affine equivariant in the sense that
		\[
		\bo T(\bo A\bo x + \bo b) = \bo A \bo T(\bo x) +\bo b,
		\]
		for all full rank $p \times p$ matrices $\bo A$ and all $p$-variate vectors $\bo b$.
	\end{definition}
	
	Thus, location and scatter functionals are a way to describe centrality and spread of the data and are then estimated by replacing $F_{\bo x}$ with the empirical distribution.
	Probably the most widely used pair of location and scatter functionals are the expected value $\E(\bo x)$ and the covariance matrix $\cov(\bo x)$.
	
	The literature is however full of many alternatives which have different desirable properties, like robustness or efficiency, at specific models. A large family of functionals which we will use in the following are the  $M$-estimators of location and scatter and are for example reviewed in \cite{DumbgenPaulySchweizer2015}.
	
	\begin{definition}
		$M$-functionals of location and scatter are defined by the two following implicit equations:
		\[
		\bo T(\bo x) = \E(w_1(r))^{-1} \E(w_1(r)\bo x)
		\]
		and
		\[
		\bo S(\bo x) = \E\left(w_2(r)\left(\bo x - \bo T(\bo x)\right) \left(\bo x - \bo T(\bo x)\right)^\top \right),
		\]
		where $w_1(r)$ and $w_2(r)$ are nonnegative continuous functions of the Mahalanobis distance $r=||\bo S(\bo x)^{-1/2}(\bo x - \bo T(\bo x))||$.
	\end{definition}
	
	Thus, $M$-functionals of location and scatter are weighted variants of the mean and the covariance matrix yielding them as special cases when choosing $w_1(r)=w_2(r)=1$.
	Usually the weight functions are chosen to be non-increasing to obtain estimators that may be robust. Some popular members of the family of $M$-estimators have the following weight functions
	
	\begin{itemize}
		\item Huber's $M$-estimators \cite{huber1964}
		\[
		w_1(r) = \left\{\begin{array}{cc}
		1 & r \leq c\\
		c/r & r > c
		\end{array} \right.
		\quad
		\mbox{and}
		\quad
		w_2(r) = \left\{\begin{array}{cc}
		1/\sigma^2 & r \leq c\\
		c/(r^2\sigma^2) & r > c
		\end{array} \right.,
		\]
		where $\sigma^2$ is a scaling factor chosen so that $\mathrm{E}(Q w_2(\sqrt{Q}))=p$ and $c$ is a tuning constant chosen to satisfy $q = Pr(Q \leq c^2)$, where $Q\sim\chi_p^2$.
		\item $M$-estimators based on the likelihood of a $t$-distribution having $\nu \geq 1$ degrees of freedom \cite{KentTyler1991}
		\[
		w_1(r) = w_2(r) = \frac{p + \nu}{r^2+\nu}.
		\]
	\end{itemize}
	
	Traditionally, $M$-estimators of location and scatter are computed via fixed point algorithms which are iterated from an initial starting point until the difference in successive functional values is less than some predetermined threshold. Depending on the weight functions there are however also other algorithms available, see e.g. \cite{DumbgenNordhausenSchuhmacher2016}.
	
	A compromise here in the iterative process are the so called one-step $M$-estimators of location and scatter which start with a pair of location and scatter functionals and then use just one updating step to obtain weighted new functionals. A scatter functional from this family which we will consider later is the scatter matrix of fourth moments which starts with the pair ($\bo T_1,\,\bo S_1$)=($\E, \cov$) and yields eventually
	\[
	\cov_4(\bo x) = \frac{1}{p+2} \E \left(r^2 \left(\bo x - \bo T_1(\bo x)\right) \left(\bo x - \bo T_1(\bo x)\right)^\top\right),
	\]
	thus having the weight function $w_2(r) = r^2/(p+2)$, where $r=||\bo S_1(\bo x)^{-1/2}(\bo x - \bo T_1(\bo x))||$.
	
	Scatter functionals are mainly investigated in the context of elliptical distributions where it is a well-known fact that they are all proportional to each other given they exist \cite{NordhausenTyler2015}. However, as the Gaussian distribution is the only elliptical distribution with independent components, other properties of scatter functionals are of interest in NGCA.
	For example the properties of full and block independence for scatter functionals are defined in \cite{NordhausenTyler2015}.
	
	\begin{definition}
		A scatter functional $\bo S(\bo x)$ is said to have the full independence property if
		\[
		\bo S(\bo x) = \bo D(\bo x)
		\]
		for all $\bo x$ having independent components where $\bo D(\bo x)$ denotes a diagonal matrix.
		
		If the $p$-variate vector $\bo x = (\bo x_1, \ldots, \bo x_k)^\top$ has $k$ independent blocks with corresponding block dimensions $p_1,\ldots,p_k$, then a scatter functional $\bo S(\bo x)$ is said to have the block independence property if
		\[
		\bo S(\bo x) = \bo B(\bo x),
		\]
		where $\bo B(\bo x)$ is the block diagonal matrix with block dimensions $p_1,\dots,p_k$.
	\end{definition}
	
	Most scatter functionals do not posses the full or block independence property, however $\cov$ and $\cov_4$ do. All scatter functionals are however diagonal and block diagonal in case when all but one of the independent parts are symmetric \cite{NordhausenTyler2015}. Exploiting the concept of symmetry, symmetrized scatter functionals can be defined.
	
	\begin{definition}
		Let $\bo S$ denote any scatter functional, then its symmetrized version is defined as
		\[
		\bo S_{sym}(\bo x) := \bo S(\bo x^1 - \bo x^2),
		\]
		where $\bo x^1$ and $\bo x^2$ are independent copies of $\bo x$.
	\end{definition}
	
	For example \cite{NordhausenTyler2015} show that every symmetrized scatter functional possess the full and block independence property. Note also that $\cov$ and $\cov_4$ can actually be expressed as functions of pairwise differences and that symmetrized scatter functionals do not require a location functional. Actually, they are usually computed using all pairwise differences and computing the original scatter with respect to the origin.
	Symmetrized $M$-estimators of scatter are investigated in \cite{SirkiaTaskinenOja2007}, while the computational issues are especially discussed in \cite{DumbgenNordhausenSchuhmacher2016,MiettinenNordhausenTaskinenTyler2016}.
	
	\section{NGCA and ICA}
	\label{Radojicic:sec:3}
	
	The non-Gaussian component analysis (NGCA) model we will consider in the following is defined as follows.
	
	\begin{definition}
		A (centered) $p$-variate vector $\bo x$ follows the NGCA model  if it can be decomposed as
		\[
		\bo x = \bo A \bo z = \bo A_1 \bo s + \bo A_2 \bo n,
		\]
		where $\bo z = (\bo s^\top \bo n^\top)^\top$ is a latent $p$-variate vector consisting of the $q$-variate non-Gaussian signal vector $\bo s$ and the $(p-q)$-variate Gaussian noise vector $\bo n$. The signal and noise vectors are independent and locations and scales are fixed using a pair of location and scatter functionals as $\bo T(\bo z) = \bo 0$ and $\bo S(\bo z) = \bo I_p$, where $\bo S$. The full-rank $p \times p$ matrix $\bo A$ is called the mixing matrix and $\bo A_1$ and $\bo A_2$ are $p \times q$ and $p \times (p-q)$ matrices with ranks $q$ and $p-q$ respectively and specify the signal and noise parts of $\bo x$.
	\end{definition}
	
	The signal dimension $q$ is the largest value separating between the signal and noise values. That is, there exists no $q$-variate vector $\bo a$ such that $\bo a^\top \bo s$ has a normal distribution, and also, $q$ is the largest such number ensuring that $\bo n$ is a Gaussian noise vector. Still, the two matrices $\bo A_1$ and $\bo A_2$ are not identifiable as both can be post-multiplied by $q\times q$ and $(p-q)\times (p-q)$ dimensional orthogonal matrices respectively and consequentially $\bo A$ is not identifiable either.
	
	The goal of non-Gaussian component analysis is thus to find a $p \times p$ full rank unmixing block matrix
$$
\bo W = (\bo W_1^\top \bo W_2^\top)^\top=\begin{pmatrix}
	                                                                                                                                                 \bo W_1 \\
	                                                                                                                                                 \bo W_2
	                                                                                                                                               \end{pmatrix},
$$
with submatrices $\bo W_1$ and $\bo W_2$, such that $\bo W_1 \bo x$ recovers the non-Gaussian signal subspace and $\bo W_2 \bo x$ the Gaussian noise subspace.\\

There are also several closely related models which we would like to introduce.
	
	The independent component analysis (ICA) model can be seen as an extreme case of the NGCA model where all components of $\bo s$ are independent and $q$ is either $p-1$ or $p$. In that case $\bo A$ is identifiable up to the order and the signs of its rows, and therefore, in this case, one can think of $\bo W$ as its inverse, keeping in mind that it is well defined up to the order and the signs of its rows. ICA is for example widely used in the analysis of biomedical signals and has many other applications; for details see for example \cite{comon2010handbook,NordhausenOja2018}.
	
	A compromise between NGCA and ICA is the non-Gaussian independent component model (NGICA) which is an NGCA model where all components of $\bo s$ are independent and the ICA model is thus a special case. The NGICA model has the advantage over the general NGCA model that the signal components of $\bo s$ are identifiable up to their order and signs. NGICA was for example considered in \cite{NordhausenOjaTylerVirta2017,RiskMattesonRuppert2019,JinRiskMatteson2019}.
	
	NGCA on the other hand can be seen as a special case of independent subspace analysis (ISA), where it is assumed that the latent vector $\bo z$ consists of $k$ independent blocks and these subspaces need to be identified. For details about ISA see for example \cite{Theis2006,NordhausenOja2011}.
	
	As mentioned above, there are many methods to estimate the unmixing matrix in NGCA where many of them are based on projection pursuit ideas. The approach of interest in this paper is based however on the simultaneous use of two scatter functionals $\bo S_1$ and $\bo S_2$.
	
	In the beginning we choose $\bo S_1 = \cov$ and $\bo S_2 = \cov_4$ and define the fourth-order-blind-identification (FOBI) functional as:
	
	\begin{definition}\label{FOBI}
		Let $\bo x$ be a $p$-variate random vector with finite fourth moments and set $\bo S_1 = \cov$ and $\bo S_2 = \cov_4$. Then the FOBI functional is defined as the $p \times p$ matrix-valued functional $\bo W$ which simultaneously diagonalizes $\bo S_1$ and $\bo S_2$. That means
		\[
		\bo W(\bo x) \bo S_1(\bo x) \bo W(\bo x)^\top = \bo I_p \quad \mbox{and} \quad
		\bo W(\bo x) \bo S_2(\bo x) \bo W(\bo x)^\top = \bo D(\bo x),
		\]
		where $\bo D(\bo x)$ is a diagonal matrix with decreasing diagonal elements.
	\end{definition}
	
	For convenience and when the context is clear the dependence on $\bo x$ of $\bo S_1$, $\bo S_2$, $\bo W$ and $\bo D$ will be omitted. The FOBI functional $\bo W$ is usually obtained by first whitening $\displaystyle \bo x \mapsto \bo x^{st} = \bo S_1(\bo x)^{-1/2}(\bo x - \E(\bo x))$ and then performing an eigenvalue-eigenvector decomposition of $\bo S_2( \bo x^{st}) = \bo U \bo D \bo U^\top$. It can then be shown that $\bo W = \bo U \bo S_1^{-1/2}$, and that $\bo D$ in the eigenvalue-eigenvector decomposition of $\bo S_2( \bo x^{st})$ is the same $\bo D$ from the
	Definition \ref{FOBI} of the FOBI functional. The latent components $z_1,\ldots, z_p$ are then obtained as $\bo z = \bo W \bo x$. The intuition behind $\bo W = \bo U \bo S_1^{-1/2}$ is that $\bo W = \bo U \bo S_1^{-1/2}$ gives latent components $\bo z = \bo W \bo x$ obtained by first whitening $\bo x$ with respect to $\bo S_1$ and then choosing $\bo z$ to be the principal components, with respect to $\bo S_2$ of the whitened $\bo x$.

In \cite{miettinen2014fourth} it is shown that in the ICA model the diagonal elements of $\bo D$, $d_1,\ldots, d_p$ correspond to kurtosis measures of latent variables $\bo z$ yielding $d_i=1$ if and only if $\E (z_i^4)=3$. Thus, in ICA, the FOBI functional is well-defined (up to signs) if all independent components have distinct kurtoses and in that case $\bo z$ corresponds to the original independent components up to signs and order.
	
	FOBI was originally suggested as an ICA method in \cite{cardoso1989source} and considered in an exploratory data analysis context in \cite{Nordhausen:2011}, and for NGCA and NGICA for example in \cite{NordhausenOjaTylerVirta2017}, while recently reviewed in \cite{NordhausenVirta2019}.\\

	Recently it was discovered that not only the combination $\cov$ and $\cov_4$ is useful but that in general
	\[
	\bo W(\bo x) \bo S_1(\bo x) \bo W(\bo x)^\top = \bo I_p \quad \mbox{and} \quad
	\bo W(\bo x) \bo S_2(\bo x) \bo W(\bo x)^\top = \bo D(\bo x),
	\]
	is of interest outside of an elliptical model where $\bo S_1$ and $\bo S_2$ can be arbitrary scatter functionals or are sometimes required to satisfy certain properties. The reason why the combination $\bo S_1-\bo S_2$ is considered especially outside an elliptical model is that if $\bo x$ has an elliptical distribution all scatters calculated at $\bo x$, provided that they exist, are proportional to each other.

In \cite{Oja:2006,Nordhausen_Oja_Ollila_2016} it is shown that any two scatter functionals which have the full independence property can be used to as an ICA method. The approach as a general exploratory method was introduced as invariant coordinate selection (ICS) \cite{TylerCritchleyDumbgenOja:2009} and useful for example for finding groups or outliers and as a transformation-retransformation method in multivariate nonparametrics \cite{TylerCritchleyDumbgenOja:2009,ics,ArchimbaudNordhausenRuizGazen2018}. For the exploratory use, there are also some guidelines provided in \cite{TylerCritchleyDumbgenOja:2009} on how to choose the two scatters while arguing that there is no general best combination.\\
	
	For two squared dispersion measures $S_1$ and $S_2$, one can define a generalized kurtosis measure with respect to $S_1$--$S_2$ as $\mathrm{Ku}(x)=S_2(x)/S_1(x)$. Furthermore, for scatter functional $\bo S$ and random vector $\bo z=(z_1,\dots,z_p)$,  $S(z_i):=\bo e_i^\top \bo S(\bo z)\bo e_i=\bo S(\bo z)_{ii}$ is a squared dispersion measure for every $i=1,\dots,p$, where $\bo e_i$ is the $i$-th vector of canonical bases of $\mathbb{R}^p$. In that manner, for two scatters $\bo S_1$ and $\bo S_2$, and a latent vector $\bo z=(z_1,\dots,z_p)$, $\bo S_2(\bo z)_{ii}/\bo S_1(\bo z)_{ii}$ can be interpreted as generalized kurtosis measures for the corresponding latent component $z_i$, with respect to $\bo S_1$--$\bo S_2$, for every $i=1,\dots,p$. Relevant for our purpose is that for any combination, $\bo S_1$ and $\bo S_2$, of scatter functionals and for any vector $\bo u\in\mathbb{R}^p$, the diagonal elements $d_1,\ldots, d_p$ of $\bo D$ satisfy,
$$
\frac{\bo u^\top\bo S_2(\bo z)\bo u}{\bo u^\top\bo S_1(\bo z)\bo u}=\sum_{i=1}^{p}u_i^2d_i.
$$
Therefore, for each $i$, $d_i=\bo S_2(\bo z)_{ii}/\bo S_1(\bo z)_{ii}$, gives the marginal kurtosis of $z_i$ with respect to $\bo S_1$--$\bo S_2$. In that manner, standard kurtosis can be considered a kurtosis measure with respect to $\cov$--$\cov_4$.

	In the following we will give results on how to use other scatter functionals besides the FOBI combination for NGCA and NGICA. Prior to stating any formal results we will introduce the following ordering. Let  $(d_1,\dots,d_p)$ be the vector in $\mathbb{R}^p$ such that $p-q$ of its components are all equal and the rest, $q$ of them, mutually distinct and distinct from the $p-q$ equal ones. We say that it is ordered in \textit{decreasing-to-equal} order if $d_1>d_2>\cdots>d_q$ and $d_{q+1}=\cdots=d_p$.\\

As the basic NGCA model has two independent blocks where at least the noise block is symmetric, basically any two scatter functionals can be used for this purpose.

\begin{result}\label{res1}
Let $\bo x$ follow an NGCA model formulated using location functional $\bo T$ and scatter functional $\bo S_1$ and let $\bo S_2$ be a scatter functional different from $\bo S_1$. Write $\bo W={\bo U}^\top\bo S_1(\bo x)^{-1/2}$, where $\bo U$  is the matrix of unit eigenvectors of $\bo S_2\left(\bo S_1^{-1/2}(\bo x-\bo T(\bo x))\right)$ (with corresponding eigenvalues in \textit{decreasing-to-equal} order). If there exists no such $q$-variate vector $\bo u$ with $\bo u^\top\bo u=1$ such that $\bo u^\top \bo s$ has the same kurtosis in the $\bo S_1$--$\bo S_2$ sense as a Gaussian component and if all non-Gaussian components $\bo s$ have mutually distinct kurtoses in $\bo S_1$--$\bo S_2$ sense, then
$$
\bo W\bo x=((\bo O_1\bo s)^\top\,(\bo O_2\bo n)^\top)^\top,
$$
where $\bo O_1,\,\bo O_2$ are orthogonal matrices.

\end{result}
	
	There should be $p-q$ equal elements in $\bo D$ which give the directions for the Gaussian subspace, however the specific value which corresponds to a Gaussian component might depend on $\bo S_1$, $\bo S_2$ and $\bo s$ and might therefore be difficult to identify in a finite data setting. Also as in general in NGCA, only the subspaces can be identified. Making the stronger assumption of an NGICA model helps in this case, but the chosen scatters are then required to have the block independence property.

\begin{result}\label{res2}
Let $\bo x$ follow an NGICA model formulated using location functional $\bo T$ and scatter functional $\bo S_1$ and let $\bo S_2$ be a scatter functional different from $\bo S_1$, where $\bo S_1$ and $\bo S_2$ have the block-independence property. Write $\bo W={\bo U}^\top\bo S_1(\bo x)^{-1/2}$, where $\bo U$  is the matrix of unit eigenvectors of $\bo S_2\left(\bo S_1^{-1/2}(\bo x-\bo T(\bo x))\right)$ (with corresponding eigenvalues in \textit{decreasing-to-equal} order). If there exists no such $q$-variate vector $\bo u$ with $\bo u^\top\bo u=1$ such that $\bo u^\top \bo s$ has the same kurtosis in the $\bo S_1$--$\bo S_2$ sense as a Gaussian component and if all non-Gaussian components $\bo s$ have mutually distinct kurtoses in $\bo S_1$--$\bo S_2$ sense, then
$$
\bo W\bo x=((\bo J\bo s)^\top\,(\bo O\bo n)^\top)^\top,
$$
where $\bo J$ is a diagonal matrix with diagonal elements $1,\,-1$ and $\bo O$ is an orthogonal matrix.

\end{result}
		
The requirement of block independence property can be relaxed under certain circumstances.

\begin{result}\label{res3}
Let $\bo x$ follow an NGICA model formulated using location functional $\bo T$ and scatter functional $\bo S_1$ such that all but one component of $\bo s$ are symmetric and let $\bo S_2$ be a scatter functional different from $\bo S_1$. Write $\bo W=\bo U^\top\bo S_1(\bo x)^{-1/2}$, where $\bo U$  is the matrix of unit eigenvectors of $\bo S_2\left(\bo S_1(\bo x)\right)$ (with corresponding eigenvalues in \textit{decreasing-to-equal} order). If there exists no such $q$-variate vector $\bo u$ with $\bo u^\top\bo u=1$ such that $\bo u^\top \bo s$ has the same kurtosis in the $\bo S_1$--$\bo S_2$ sense as a Gaussian component and if all non-Gaussian components $\bo s$ have mutually distinct kurtoses in $\bo S_1$--$\bo S_2$ sense, then
$$
\bo W\bo x=((\bo J\bo s)^\top\,(\bo O\bo n)^\top)^\top,
$$
where $\bo J$ is a diagonal matrix with diagonal elements $1,\,-1$ and $\bo O$ is an orthogonal matrix.

\end{result}

To conclude this section we would, however, like to point out that in NGCA and NGICA the Gaussian subspace can still be separated from the non-Gaussian subspace if the kurtoses in $\bo S_1$--$\bo S_2$ sense of the signals are not distinct as long as they differ from the corresponding Gaussian value.

	\section{Testing the signal dimension in NGCA and NGICA}
	\label{Radojicic:sec:4}
	
	FOBI is such a popular functional since it is solely moment based and therefore analytical considerations are fairly easy. However, it requires strong moment assumptions and suffers from a lack of robustness. In the NGCA and NGICA context the FOBI functional has the advantage that the values in $\bo D$ of Gaussian components are known to be one. Therefore, in these models, in \cite{NordhausenOjaTylerVirta2017,Nordhausen:2016} is suggested the testing procedure to test the hypothesis
	\[
	H_{0k}: \mbox{There are exactly} \ k \ \mbox{non-Gaussian components}
	\]
by testing that there are $p-k$ eigenvalues in $\bo D$ equal to 1.
	
	The criterion used in \cite{NordhausenOjaTylerVirta2017,Nordhausen:2016}, to identify the eigenvalues which are closest to $1$,  is $(d_i-1)^2$, thus the variance of the $p-k$ elements of $\bo D$ closest to $1$ is used as  the test statistic.
	Denote $d_{(i)}$, $i=1,\ldots,p$ the ascending ordered eigenvalues in the sense above, then the test statistic from \cite{NordhausenOjaTylerVirta2017,Nordhausen:2016} for a sample $\bo x_1,\ldots, \bo x_n$ is
	\[
	T_k = n \sum_{i=1}^{p-k} \left(d_{(i)}-1 \right)^2.
	\]

	In \cite{NordhausenOjaTylerVirta2017,Nordhausen:2016} it is then shown that assuming $\mathrm{E}(z_i^4)$ exist for $i=1,\ldots,p$ and that there is no $q$-variate vector $\bo u$ with $\bo u^\top\bo u=1$ such that $\mathrm{E}((\bo u^\top \bo s)^4) = 3$, where $\bo s$ is the signal component, one can use FOBI for estimating the signal and noise subspaces in NGCA and NGICA models as well as making inference about their dimensions.

Before stating the result that gives the limiting distribution of the test statistic $T_k$ and enables for testing of $H_{0k}$, $k\in \{1,\ldots,p\}$, we define $\bo U_k$ to be the $p\times k$ matrix of eigenvectors of $\bo S_2$ that correspond to the aforementioned  $p-k$ eigenvalues in $\bo D$ that are closest to $1$, and the statistic $\displaystyle T_k^*=n\,\mathrm{tr}(((\bo 0, \bo I_{p-k})\bo U_k(\bo S_2 - \bo I_p)\bo U_k^\top(\bo 0, \bo I_{p-k})^\top)^2)$. The statistic $T_k^*$ then corresponds to the test statistic for testing $H_{0k}$ in case where the noise part is known.
	
	\begin{result}\label{th1}
		Under the previously stated assumptions and under $H_{0q}$
		\begin{enumerate}
			\item for $k < q$, $(p+2)^2T_k \rightarrow_P c$ for some $c>0$  as $n\rightarrow\infty$,
			\item for $k = q$, $(p+2)^2T_k \rightarrow_d C_k$  as $n\rightarrow\infty$ and
			\item for $k > q$, $(p+2)^2T_k \leq (p+2)^2T_k^* \rightarrow_d C_k$, as $n\rightarrow\infty$,
		\end{enumerate}
		
		where
		$$
		C_k\sim 2\sigma_1 Q_1+(2\sigma_1+\sigma_2(p-k))Q_2,
		$$
		where $Q_1$, $Q_2$ are independent, chi-squared distributed, random variables with $(p-k-1)(p-k+2)/2$ and $1$ degrees of freedom respectively, and $\sigma_1^2=Var(||\bo z||^2)+8$, $\sigma_2=4$.
		
	\end{result}
	
	The proof of the Result \ref{th1} can be found in \cite{NordhausenOjaTylerVirta2017}. In this setting, the null hypothesis is rejected if $T_k\geq c_{k,\alpha}$, where $c_{k,\alpha}$ is chosen so that $\mathrm{P} (C_k\geq c_{k,\alpha})=\alpha$. Note that, in order to find $c_{k,\alpha}$ one must consistently estimate $\sigma_1$. If we write $\hat{ \bo z}_i=\hat{ \bo W} (\bo x_i - \bar{\bo x})$ $i=1,\ldots,n$, then in the NGICA model we have $\displaystyle\sigma_1=\sum_{k=1}^{p}\mathrm{E}(z_k^4)-p+8$, with a consistent estimate
	\[\displaystyle\hat \sigma_1=\frac{1}{n}\sum_{i=1}^{n}\sum_{k=1}^{p}(\hat z_i)_k^4-p+8
	.\]

In the wider NGCA model $\sigma_1$ can be consistently estimated by
 \[
\hat{\sigma}_1=\tfrac{1}{n}\sum_{i=1}^{n}||\hat z_i||^4-p^2+8.\]
	
	Besides $T_k$ \cite{NordhausenOjaTylerVirta2017} proposes also alternative for this problem such as
	$$
	\frac{(p+2)^2T_{k,1}}{2\hat{\sigma}_1^2}\quad \text{and}\quad\frac{(p+2)^2T_{k,2}}{2\hat{\sigma}_1^2+4(p-k)},
	$$
	where $T_{k,1}= n \left(\sum_{i=1}^{p-k} d_{(i)}^2-\left(\sum_{i=1}^{p-k} d_{(i)} \right)^2\right)$ and $T_{k,2}=n \left(\sum_{i=1}^{p-k} (d_{(i)}-1)\right)^2$. Under the true $H_{0k}$, proposed test statistics have chi-squared distributions with $(p-k-1)(p+2-k)/2$ and $1$ degrees of freedom respectively. One can show that $\displaystyle T_{k,1}+T_{k,2}\sim\chi_{(p-k-1)(p+2-k)/2+1}^2$, and argue that $T_{k,1}$ provides a test statistic for testing the equality of $p-k$ eigenvalues closest to $1$, while $T_{k,2}$ measures the deviation of the mean of those eigenvalue from the theoretical value of one. In \cite{NordhausenOjaTylerVirta2017} it is also argued that those two statistics use less information than $T_k$, and are therefore in most cases less powerful and that the limiting behaviour of their sum is quite similar to the one of $T_k$.\\
	
	Result \ref{th1} gives the limiting distribution of $T_k$, and therefore when using it in practice, due to the involvement of higher order moments, one might need very large sample sizes for the result to hold. For the case of small sample sizes, in \cite{NordhausenOjaTylerVirta2017} is proposed to estimate the distribution of the test statistic under the null by bootstrapping samples from distribution for which the null hypothesis $H_{0k}$ is true and which is as close as possible to the empirical distribution of observed sample.

Let $\bo X = (\bo x_1,\ldots,\bo x_n)$ be a data sample, and let $\bar{\bo x}$ denote the sample mean vector. Further, let  $\displaystyle\hat{\bo W}=(\hat{\bo W}_1^\top\, \hat{\bo W}_2^\top)^\top$ be the sample estimates of the FOBI unmixing matrices where the partition $\displaystyle(\hat{\bo W}_1^\top\, \hat{\bo W}_2^\top)^\top$ was done according to the descending order of the eigenvalues in $\hat{\bo D}$ in sense as described in Section~\ref{Radojicic:sec:3}. Furthermore, let
$\displaystyle
\hat{\bo S}=(\hat{\bo s}_1,\ldots,\hat{\bo s}_n)=\hat{\bo W}_1 (\bo X- \bar{\bo x} \bo 1_n^\top) \in\mathbb{R}^{k\times n}$ and $\displaystyle\hat{\bo N}=(\hat{\bo n}_1,\ldots,\hat{\bo n}_n)=\hat{\bo W}_2(\bo X- \bar{\bo x} \bo 1_n^\top)\in\mathbb{R}^{(p-k)\times n}$
be the matrices of  the estimated signal and noise vectors, $\hat {\bo s}_i$ and $\hat{\bo  n}_i$, $i\in 1,\ldots, n$ respectively. $\bo 1_n$ denotes here an $n$-vector full of ones. The proposed strategy in the NGICA model is using non-parametric bootstrap to create matrices $\bo S^*\in\mathbb{R}^{k\times n}$ by componentwise(row-wise)-independently sampling with replacement from $\hat{\bo S}$, and using parametric bootstrap to create $\bo N^*\in \mathbb{R}^{(p-k)\times n}$ as a random sample from $N(\bo 0,\bo I_{p-k})$. Resulting bootstrap sample is then
$\bo X^*=\hat{\bo W}^{-1}({\bo S^*}^\top\,{\bo N^*}^\top)^\top$.

A similar approach for NGCA model was introduced in \cite{Nordhausen:2016}. The strategy is to initially sample with replacement an $n$-dimensional sample $\tilde{\bo X}\in\mathbb{R}^{p\times n}$ from $\bo X$ and then estimate its signal matrix $\bo S^*=\hat{\bo W}_1 \tilde{\bo X}$. In order for the noise space to be Gaussian transform $\tilde{\bo X}$ into $\bo X^*=\hat{\bo W}^{-1}({\bo S^*}^\top\,{\bo N^*}^\top)^\top$, where
$\bo N^*\in \mathbb{R}^{(p-k)\times n}$ is an $n$-dimensional random sample from $N(\bo 0,\bo I_{p-k})$.\\

We showed earlier that using the general two scatter functionals approach is possible for NGCA and NGICA given the scatter functionals fulfill certain properties. However it is already not in general possible to say which eigenvalues correspond to directions indicating the Gaussian subspace. Thus deriving a general asymptotic test for any scatter combination does not sound feasible. However the bootstrap strategy described above for FOBI can be adapted.\\

One of the alternative test statistics mentioned earlier which considers only the variance of the eigenvalues can be used here, when adding the additional assumption that the Gaussian subspace is larger than any set of the signal subspaces which would share the same eigenvalue, which is for example in NGICA anyway required.

Hence, for $k\in\{ 0,\ldots,p-2$\} one can test $H_{0k}$ by examining the variance of the $p-k$ eigenvalues closest together in that sense. In that manner we propose a bootstrap procedure that uses two scatter matrices $\bo S_1$, $\bo S_2$ and a location functional $\bo T$ and starts with a sample $\bo X=(\bo x_1,\ldots,\bo x_n)\in\mathbb{R}^{p\times n}$. Using sample estimators $\hat{\bo T}$ and $\hat{\bo S}_1$ of $\bo T$ and $\bo S_1$ respectively, scatter estimator $\bo S_2$ as its sample estimate based on standardized sample $\hat{\bo S}_1^{-1/2}(\bo X-\hat{\bo T}\bo 1_n^\top)$, and calculates corresponding unmixing matrix $\hat{\bo W}$ as discussed in Section~\ref{Radojicic:sec:3}. The test statistic used for testing $H_{0k}$ is then

	\[
	\hat{t}_{k} = n \sum_{i=1}^{p-k} \left(d_{(i)}-\frac{1}{p-k} \sum_{j=1}^{p-k} d_{(j)} \right)^2,
	\]

where $d_{(1)},\ldots, d_{(p-k)}$ are those $p-k$ eigenvalues of $\hat{\bo S}$ contained in $\hat{\bo D}$ that have the smallest variance of all $p-k$ - subsets of the set of eigenvalues in $\hat{\bo D}$. Thus, $\hat{t}_{k}$ is the estimator of the variance of those $p-k$ eigenvalues of $\hat{\bo S}_2$ that correspond to the Gaussian components.
	
Once the eigenvalues corresponding to the signal and noise space have been identified one can order the diagonal elements of $\hat{\bo D}$ in a way that the last $p-k$ eigenvalues form a $p-k$ - subset of set of all eigenvalues of $\hat{\bo S}_2$ with the minimal variance, and obtain the corresponding partitioning of $\displaystyle\hat{\bo W}=(\hat{\bo W}_1^\top\, \hat{\bo W}_2^\top)^\top$. Finally, the signal and the noise parts of the latent sample $\bo Z$ are estimated as $\hat{\bo s}_i=\hat{\bo W}_1(\bo x_i - \hat{\bo T})$ and $\hat{\bo n}_i={\hat{\bo W}_2}(\bo x_i - \hat{\bo T})$ respectively yielding the matrices $\hat{\bo S}\in\mathbb{R}^{k\times n}$ and $\hat{\bo N}\in\mathbb{R}^{(p-k)\times n}$ which collect the estimated signal and noise vectors.\\

Since the bootstrapping strategy for the signal part is dependent on the model, in the NGCA model we use the non-parametric bootstrap to create  the signal sample $\bo S^*$ by sampling with replacement from $\hat{\bo S}$.

In the NGICA model, where signal components are mutually independent, we use non-parametric bootstrap to create matrix  $\bo S^*\in\mathbb{R}^{k\times n}$ by componentwise(row-wise)-independently sampling with replacement from $\hat{\bo S}$.\\

We also propose two strategies for sampling the noise component. Parametric bootstrap creates noise sample $\bo N^*\in \mathbb{R}^{(p-k)\times n}$ as a random sample from $N(\bo 0,\cov(\bo N))$, while the nonparametric bootstrap creates noise sample $\bo N^*=(\bo n_1^*,\ldots,\bo n_n^*)\in \mathbb{R}^{(p-k)\times n}$, such that  $\bo{n}_i^* \leftarrow \bo O_i \hat{\bo n}_i$, $i=1,\ldots,n$, where $\bo O_i$ is a random orthogonal $p-k \times p-k$ matrix. The nonparametric strategy does not directly target a normal noise but  assumes spherical noise as a proxy.

For the latent component sample $\bo Z^*=({\bo S^*}^\top\, {\bo N^*}^\top)^\top$ obtained by bootstrapping procedure explained above, set $\bo X^*=\bo W^{-1}\bo Z^*$. Finally, assuming that $\bo X_1^*,\ldots,\bo X_M^*$ are $M$ independent bootstrap samples obtained as described above and $\hat{t}^*_{i,k}=\hat{t}_k(\bo X_i^*)$ are the corresponding test statistics, the bootstrap $p$-value is given by
$$
\hat{p}=\frac{\#(\hat{t}^*_{i,k} \geq \hat{t}_{k}) + 1}{M + 1}.
$$

The bootstrapping procedure for the combination of any two scatters is given in a schematic view in Algorithm~\ref{alg::boot}.

\begin{algorithm}[htb]\label{alg::boot}
		Set the proposed dimension $k$;\;\\
		Set the number of bootstrap samples $M$;\;\\
		Choose two scatter functionals $\bo S_1$ and $\bo S_2$ and location functional $\bo T$;\;\\
		Starting with the observed sample $\bo X=(\bo x_1,\ldots,\bo x_n)$, $\bo x_i\in\mathbb{R}^p$ estimate $\hat{\bo T}=\bo T(\bo X)$, $\hat{\bo S}_1=\bo S_1(\bo X)$;\;\\
        Calculate centered and standardized sample $\bo X^c=(\bo x_1^c,\ldots,\bo x_n^c)$ and $\bo X^{st}=(\bo x_1^{st},\ldots,\bo x_n^{st})$ respectively, where $\bo x_i^{c}=\bo x_i-\hat{\bo T}$, $\bo x_i^{st}=\hat{\bo S}_1^{-1/2}(\bo x_i-\hat{\bo T})$, $i=1,\ldots,n$;\;\\
		Estimate $\hat{\bo S}_2=\bo S_2(\bo X^{st})$ and calculate its eigenvalue-eigenvector decomposition $\hat{\bo S}_2=\hat{\bo U}\hat{\bo D}{\hat{\bo U}}^\top$;\;\\
		Calculate two-scatter functional $\hat{\bo W}=$
		$\hat{\bo U}{\hat{\bo S}_1}^{-1/2}$;\;\\
		Order eigenvalues in $\hat{\bo D}$ so that the variance of the last $p-k$ eigenvalues in $\hat{\bo D}$ is minimal and derive the corresponding partitioning of $\hat{\bo W}=(\hat{\bo W}_1^\top\,\hat{\bo W}_2^\top)^\top$;\;\\
		Compute the test statistic $\displaystyle\hat{t}_{k}=n \sum_{i=1}^{p-k} \left(d_{i}-\frac{1}{p-k} \sum_{j=1}^{p-k} d_{j} \right)^2$ as the estimate of the variance of the last $p-k$ eigenvalues in $\hat{\bo D}$;\;\\
		Calculate the signal estimate $\hat{\bo S}=(\hat{\bo s}_1,\ldots,\hat{\bo s}_n)=\hat{\bo W}_1\bo X^{c}$ and the noise estimate $\hat{\bo N}=(\hat{\bo n}_1,\ldots,\hat{\bo n}_n)=\hat{\bo W}_2\bo X^{c}$;\;\\
		Choose a bootstrapping strategy for the noise;\;\\
        Choose the model suitable for the signal;\;\\
		\For{$j \in \{ 1, \ldots , M \}$}{
			\If{Strategy = parametric bootstrap}{$\bo{n}_i^*\leftarrow N_{p-k}(\bo 0, \cov(\hat{\bo N}))$, $i=1,\ldots,n$;\;}
			\If{Strategy = nonparametric bootstrap}{$\bo{n}_i^* \leftarrow \bo O_i \hat{\bo n}_i$, $i=1,\ldots,n$, where $\bo O_i$ is a random orthogonal $p-k \times p-k$ matrix;\;}
			\If{Model = NGCA}{Sample $\bo S^*$ with replacement from $\hat{\bo S}$;\;}
			\If{Model = NGICA}{For each $j=1,\dots, k$ sample with replacement $j$-th signal component $(s_{j,1}^*,\ldots, s_{j,n}^*) \leftarrow (\hat{s}_{j,1},\ldots,\hat{s}_{j,n})$, and set $\bo S^*=[s_{i,j}^*]$;\;}
			Compute $\bo X^* = {\hat{\bo W}^{-1}}({\bo S^*}^\top\, 
			{\bo N^*}^\top)^\top$;\; \\
			Compute $\hat{t}_{j,k}^{*}$ based on $\bo X^{*}$;\;}
		Return bootstrap $p$-value: $\displaystyle \hat{p}_k=[\#(\hat{t}_{j,k}^{*} \geq \hat{t}_{k}) + 1]/(M + 1)$\;
		\caption{Algorithm for testing $H_{0k} : q=k$.} 
\end{algorithm}
	
	
\section{Performance evaluation of the test}
	\label{Radojicic:sec:5}
	The following simulation study is performed using R 3.6.1 \cite{R} with the packages
	SpatialNP \cite{SpatialNP}, ICtest \cite{ICtest}, JADE \cite{MiettinenNordhausenTaskinen2017}, ICS \cite{ics}, png \cite{png}, RcppRoll \cite{RcppRoll} and extraDistr \cite{extraDistr}, and it was conducted to compare the  bootstrap FOBI test from \cite{NordhausenOjaTylerVirta2017} to four different testing procedures based on Algorithm \ref{alg::boot} with the expectation as the location functional and the following pairs of scatter matrices:
	\begin{enumerate}
		\item \textit{Cov - Cov4}: $\bo S_1=\cov$, $\bo S_2=\cov_4$.
\end{enumerate}

Note that there is a difference between the ``FOBI'' and the $\textit{Cov-Cov4}$ testing procedures. In the ``FOBI'' denoted case the information that the noise eigenvalues should be one is used while in the $\textit{Cov-Cov4}$ denoted case Algorithm~\ref{alg::boot} is used ignoring this information.
\begin{enumerate}
	\setcounter{enumi}{1}
		\item \textit{Cau-Hub}: $\bo S_1$ is $M$-estimator based on the likelihood of the $t$-distribution with one degree of freedom ($\nu=1$), also known as the Cauchy distribution. $\bo S_2$ is an $M$-estimator based on Huber's weight function.
		\item \textit{sCau-sHub}: is the symmetrized version of the previous setting, thus a symmetrized $M$-scatter based on the Cauchy distribution and a symmetrized $M$-scatter based on Huber's weight function.
\end{enumerate}

As estimation of both scatters in  \textit{sCau-sHub} is computationally very expensive and not feasible in the large data sets we follow a suggestion  from \cite{MiettinenNordhausenTaskinenTyler2016} to base the symmetrized scatters not on all pairwise differences but only on an ``incomplete'' set which makes it much easier to compute. For details see \cite{MiettinenNordhausenTaskinenTyler2016}.

\begin{enumerate}
	\setcounter{enumi}{3}
		\item \textit{sCauI-sHubI}: is the incomplete combination of symmetrized scatters. We compute both scatters so that all observations are contained in 100 differences.
	\end{enumerate}
	For more details on the computation of all the scatters see also the documention of the R-packages SpatialNP \cite{SpatialNP} and ICS \cite{ics}.\\

Due to the computational costs and as it seems more natural to us, in all four settings always parametric bootstrap is used for the noise part.\\

\begin{figure}[ht]
		\centering
		\includegraphics[width=1\textwidth,keepaspectratio]{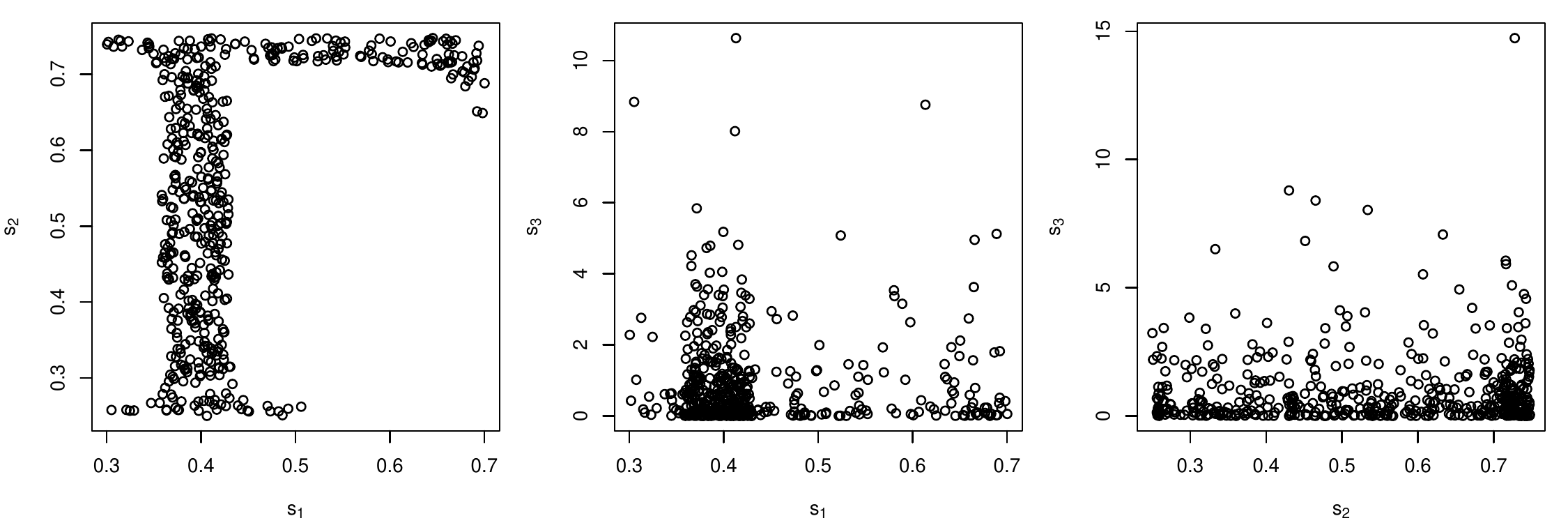}
		\caption{Scatter plots of signal components in $M1$ based on a sample of $500$.}
		\label{M1}
	\end{figure}

To compare the bootstrap tests, we consider two different settings which both are 6-variate and have each 3 signal and 3 noise components. Model $M1$ follows an NGCA model and model $M2$ an NGICA model. In all cases the $6 \times 6$ matrix $\bo A$ was simulated in each iteration independently by filling it with random $N(0,1)$ elements. The two models used are:
	
	\begin{description}
		\item[$M1$:] An NGCA model with two non-Gaussian univariate components $\bo s_1$ and $\bo s_2$, representing $x$ and $y$ axis of the Greek letter $\Gamma$ respectively, a non-Gaussian univariate component $\bo s_3$ with $\chi_1^2$ distribution and three independent Gaussian components $N(0,1)$. Hence, $p=6, \, q=3$.  Figure~\ref{M1} visualizes the three non-Gaussian components of this setting.
	\end{description}

\begin{figure}[ht]
		\centering
		\includegraphics[width=1\textwidth,keepaspectratio]{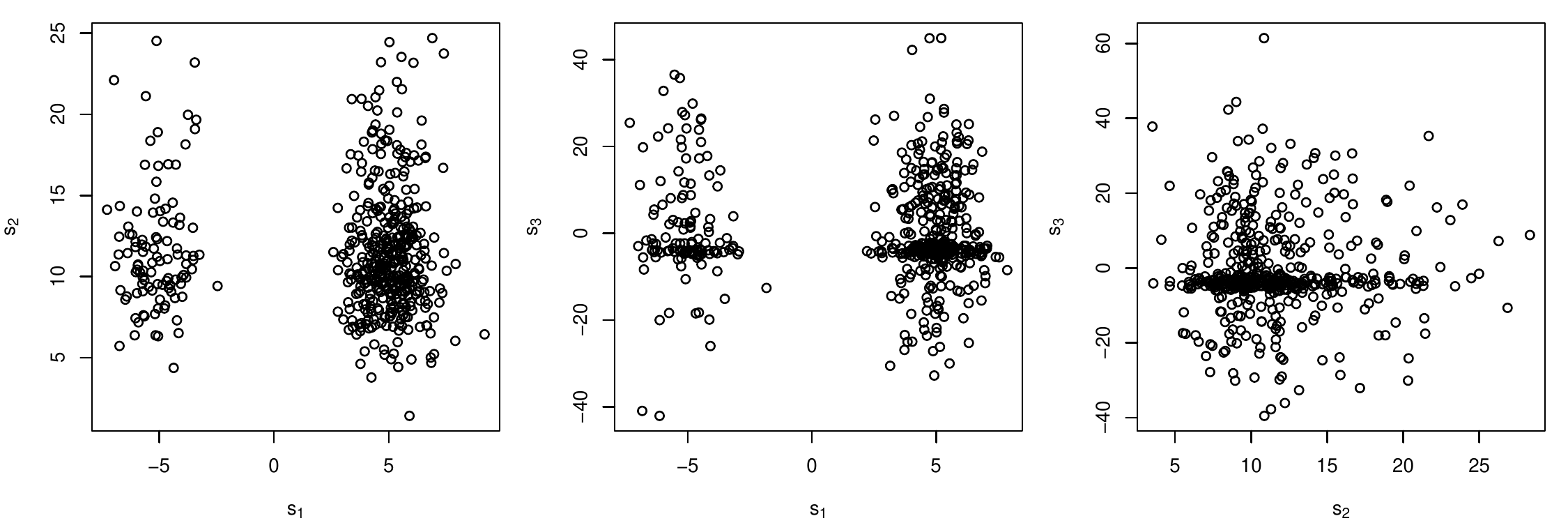}
		\caption{Scatter plots of signal components in $M2$ based on a sample of $500$.}
		\label{M2}
	\end{figure}

\begin{description}
	\item[$M2$:] An NGICA model with three independent components which all follow a Gaussian mixture model (GMM) with different parameter settings: $\displaystyle s_1 \sim  (3+\sqrt{3})^{-1}\phi_{-5,1}+(1-(3+\sqrt{3})^{-1})\phi_{5,1}$, $\displaystyle  s_2 \sim 0.7\phi_{10,2}+0.3\phi_{15,5}$ and $\displaystyle  s_3 \sim 0.4\phi_{-4,1}+0.6\phi_{2,15}$, where $\phi_{\mu,\sigma}$ denotes the pdf of the normal distribution with mean $\mu$ and variance $\sigma^2$. The three noise components are independent $N(0,1)$. Therefore, $p=6$, $q=3$. For more insight into the shape of the non-Gaussian components see Figure~\ref{M2}.
		\end{description}
	
Note that if a random variable $x$ comes from the two-component GMM, with equal variances for the components and the mixing probability is $(3+\sqrt{3})^{-1}$, then its kurtosis is equal to $3$ for all choices of means of two components. Therefore, in the model M2, the kurtosis of the component $s_1$ equals $3$. Hence, the requirements of Result~\ref{res2} are violated when the scatter combination  $\bo S_1 = \cov$ and $\bo S_2=\cov_4$ is used. Thus it is to be expected that neither \textit{Cov-Cov4} nor FOBI will be able to separate $ s_1$ form the Gaussian components, which should result in very low rejection rates in testing for $H_{02}$.\\

In order to gain insight into the robustness of the proposed testing procedures we consider also the case when in the two settings small contaminations are added. The perturbed models are denoted $M1_x$ and $M2_x$ respectively and are obtained by adding an additional perturbation (equal to $10 \ \bo 1_6$) to $0.5\%$ 
of the mixed observations.\\


For all samples $\bo X\in \mathbb{R}^{n\times p}$ from models $M1$, $M2$, $M1_x$ and $M2_x$, with sample sizes $n=500,\,1000,\,2000,\,4000$, the bootstrap $p$-values based on $M=200$ bootstrap samples were computed using the five tests described above where we use only parametric bootstrapping for the noise part. This is due to the computational complexity of the simulation and as it seems to be a more natural suggestion.  We performed all the bootstrap tests once assuming an NGCA model and once assuming an NGICA model. $1000$ repetitions where performed at the level $\alpha=0.05$ and Tables \ref{t1}-\ref{t8} report the rejection rates for $H_{02}$, $H_{03}$(true) and $H_{04}$ in all discussed settings. In the case $n= 4000$ also due to computational complexity the tests \textit{sCau-sHub} have not been performed. In our settings the non-FOBI combinations should all be able to separate the signal and noise subspaces but only the symmetrised scatters would actually be able to recover the individual signal components in model $M2$.

\begin{table}[ht]
		\centering
		\caption{$M1$ assuming NGCA model: Rejection rates in $1000$ repetitions for bootstrap tests of $H_{02}$, $H_{03}$ (true) and $H_{04}$, with $\alpha=0.05$. }
		\label{t1}
		\begin{adjustbox}{width=1\textwidth}
			\begin{tabular}{@{\extracolsep{4pt}}ccccccccccccccccccc@{}}
				\hline
				&\multicolumn{3}{c}{FOBI (boot)}&\multicolumn{3}{c}{\textit{Cov-Cov4}}&\multicolumn{3}{c}{\textit{Cau-Hub}}&\multicolumn{3}{c}{\textit{sCau-sHub}}&\multicolumn{3}{c}{\textit{sCauI-sHubI}} \\
				
				n & $k2$ & $k3$ & $k4$ & $k2$ & $k3$ & $k4$ & $k2$ & $k3$ & $k4$ & $k2$ & $k3$ & $k4$ & $k2$ & $k3$ & $k4$ \\
				\cline{2-4} \cline{5-7} \cline{8-10} \cline{11-13} \cline{14-16} \cline{17-19}
				
				500 & 0.363 & 0.035 & 0.016 & 0.360 & 0.088 & 0.031 & 0.954 & 0.061 & 0.017 & 0.788 & 0.055 & 0.027 & 0.233 & 0.091 & 0.083 \\
				1000 & 0.553 & 0.057 & 0.016 & 0.553 & 0.074 & 0.031 & 1.000 & 0.050 & 0.011 & 1.000 & 0.053 & 0.018 & 0.717 & 0.072 & 0.059 \\
				2000 & 0.839 & 0.049 & 0.015 & 0.801 & 0.065 & 0.024 & 1.000 & 0.051 & 0.012 & 1.000 & 0.044 & 0.016 & 0.994 & 0.055 & 0.045 \\
				4000 & 0.986 & 0.057 & 0.012 & 0.977 & 0.051 & 0.012 & 1.000 & 0.055 & 0.013 & & & & & 0.045 & 0.035 \\
				\hline
			\end{tabular}
		\end{adjustbox}
	\end{table}
\begin{table}[ht]
		\centering
		\caption{$M1$ assuming NGICA model: Rejection rates in $1000$ repetitions for bootstrap tests of $H_{02}$, $H_{03}$ (true) and $H_{04}$, with $\alpha=0.05$. }
		\label{t2}
		\begin{adjustbox}{width=1\textwidth}
			\begin{tabular}{@{\extracolsep{4pt}}ccccccccccccccccccc@{}}
				\hline
				&\multicolumn{3}{c}{FOBI (boot)}&\multicolumn{3}{c}{\textit{Cov-Cov4}}&\multicolumn{3}{c}{\textit{Cau-Hub}}&\multicolumn{3}{c}{\textit{sCau-sHub}}&\multicolumn{3}{c}{\textit{sCauI-sHubI}} \\
				
				n & $k2$ & $k3$ & $k4$ & $k2$ & $k3$ & $k4$ & $k2$ & $k3$ & $k4$ & $k2$ & $k3$ & $k4$ & $k2$ & $k3$ & $k4$ \\
				\cline{2-4} \cline{5-7} \cline{8-10} \cline{11-13} \cline{14-16}
				
				500  & 0.384 & 0.047 & 0.033 & 0.402 & 0.128 & 0.061 & 0.967 & 0.190 & 0.119 & 0.813 & 0.148 & 0.078 & 0.302 & 0.168 & 0.149 \\
  1000  & 0.539 & 0.043 & 0.030 & 0.555 & 0.084 & 0.043 & 1.000 & 0.114 & 0.067 & 1.000 & 0.079 & 0.049 & 0.734 & 0.161 & 0.135 \\
  2000  & 0.831 & 0.035 & 0.029 & 0.804 & 0.054 & 0.040 & 1.000 & 0.077 & 0.057 & 1.000 & 0.046 & 0.035 & 0.996 & 0.096 & 0.072 \\
  4000  & 0.985 & 0.049 & 0.030 & 0.976 & 0.039 & 0.028 & 1.000 & 0.068 & 0.049 &  &  &  &  & 0.050 & 0.052 \\
	\hline
			\end{tabular}
		\end{adjustbox}
	\end{table}	
	\begin{table}[ht]
		\centering
		\caption{$M1_x$ assuming NGCA model: Rejection rates in $1000$ repetitions for bootstrap tests of $H_{02}$, $H_{03}$ (true) and $H_{04}$, with $\alpha=0.05$. }
		\label{t3}
		\begin{adjustbox}{width=1\textwidth}
			\begin{tabular}{@{\extracolsep{4pt}}ccccccccccccccccccc@{}}
				\hline
				&\multicolumn{3}{c}{FOBI (boot)}&\multicolumn{3}{c}{\textit{Cov-Cov4}}&\multicolumn{3}{c}{\textit{Cau-Hub}}&\multicolumn{3}{c}{\textit{sCau-sHub}}&\multicolumn{3}{c}{\textit{sCauI-sHubI}} \\
				
				n & $k2$ & $k3$ & $k4$ & $k2$ & $k3$ & $k4$ & $k2$ & $k3$ & $k4$ & $k2$ & $k3$ & $k4$ & $k2$ & $k3$ & $k4$ \\
				\cline{2-4} \cline{5-7} \cline{8-10} \cline{11-13} \cline{14-16}
				
				500 & 0.299 & 0.132 & 0.034 & 0.189 & 0.109 & 0.087 & 0.952 & 0.063 & 0.019 & 0.713 & 0.095 & 0.035 & 0.262 & 0.097 & 0.086 \\
				1000 & 0.183 & 0.092 & 0.006 & 0.149 & 0.103 & 0.072 & 1.000 & 0.054 & 0.012 & 0.963 & 0.188 & 0.045 & 0.674 & 0.129 & 0.096 \\
				2000 & 0.366 & 0.102 & 0.008 & 0.273 & 0.089 & 0.058 & 1.000 & 0.047 & 0.013 & 0.996 & 0.274 & 0.045 & 0.943 & 0.173 & 0.054 \\
				4000 & 0.705 & 0.159 & 0.019 & 0.568 & 0.147 & 0.049 & 1.000 & 0.062 & 0.021 & & & & 0.996 & 0.306 & 0.074 \\
				\hline
			\end{tabular}
		\end{adjustbox}
	\end{table}
	
\begin{table}[ht]
		\centering
		\caption{$M1_x$ assuming NGICA model: Rejection rates in $1000$ repetitions for bootstrap tests of $H_{02}$, $H_{03}$ (true) and $H_{04}$, with $\alpha=0.05$. }
		\label{t4}
		\begin{adjustbox}{width=1\textwidth}
			\begin{tabular}{@{\extracolsep{4pt}}ccccccccccccccccccc@{}}
				\hline
				&\multicolumn{3}{c}{FOBI (boot)}&\multicolumn{3}{c}{\textit{Cov-Cov4}}&\multicolumn{3}{c}{\textit{Cau-Hub}}&\multicolumn{3}{c}{\textit{sCau-sHub}}&\multicolumn{3}{c}{\textit{sCauI-sHubI}} \\
				
				n & $k2$ & $k3$ & $k4$ & $k2$ & $k3$ & $k4$ & $k2$ & $k3$ & $k4$ & $k2$ & $k3$ & $k4$ & $k2$ & $k3$ & $k4$ \\
				\cline{2-4} \cline{5-7} \cline{8-10} \cline{11-13} \cline{14-16}
				
				500 & 0.320 & 0.160 & 0.099 & 0.202 & 0.231 & 0.159 & 0.966 & 0.216 & 0.129 & 0.743 & 0.229 & 0.087 & 0.271 & 0.153 & 0.116 \\
  1000  & 0.191 & 0.131 & 0.042 & 0.145 & 0.206 & 0.150 & 1.000 & 0.133 & 0.061 & 0.961 & 0.343 & 0.110 & 0.677 & 0.244 & 0.116 \\
  2000 & 0.369 & 0.135 & 0.056 & 0.279 & 0.178 & 0.120 & 1.000 & 0.079 & 0.045 & 0.997 & 0.429 & 0.116 & 0.931 & 0.358 & 0.151 \\
  4000  & 0.708 & 0.194 & 0.049 & 0.581 & 0.213 & 0.120 & 1.000 & 0.083 & 0.039 &  &  &  & 0.998 & 0.472 & 0.115 \\
				\hline
			\end{tabular}
		\end{adjustbox}
	\end{table}
\begin{table}[ht]
		\centering
		\caption{$M2$ assuming NGCA model: Rejection rates in $1000$ repetitions for bootstrap tests of $H_{02}$, $H_{03}$ (true) and $H_{04}$, with $\alpha=0.05.$ }
		\label{t5}
		\begin{adjustbox}{width=1\textwidth}
			\begin{tabular}{@{\extracolsep{4pt}}ccccccccccccccccccc@{}}
				\hline
				&\multicolumn{3}{c}{FOBI (boot)}&\multicolumn{3}{c}{\textit{Cov-Cov4}}&\multicolumn{3}{c}{\textit{Cau-Hub}}&\multicolumn{3}{c}{\textit{sCau-sHub}}&\multicolumn{3}{c}{\textit{sCauI-sHubI}} \\
				
				n & $k2$ & $k3$ & $k4$ & $k2$ & $k3$ & $k4$ & $k2$ & $k3$ & $k4$ & $k2$ & $k3$ & $k4$ & $k2$ & $k3$ & $k4$ \\
				\cline{2-4} \cline{5-7} \cline{8-10} \cline{11-13} \cline{14-16} \cline{17-19}
				
500  & 0.076 & 0.014 & 0.007 & 0.088 & 0.047 & 0.040 & 0.938 & 0.107 & 0.095 & 0.995 & 0.081 & 0.051 & 0.817 & 0.220 & 0.112 \\
  1000  & 0.076 & 0.025 & 0.009 & 0.075 & 0.021 & 0.015 & 0.999 & 0.067 & 0.055 & 1.000 & 0.076 & 0.064 & 0.993 & 0.113 & 0.091 \\
  2000 &  0.064 & 0.007 & 0.007 & 0.050 & 0.014 & 0.007 & 1.000 & 0.029 & 0.043 & 1.000 & 0.029 & 0.043 & 1.000 & 0.050 & 0.107 \\
  4000 & 0.064 & 0.016 & 0.003 & 0.052 & 0.017 & 0.010 & 1.000 & 0.052 & 0.021 &  &  &  & 1.000 & 0.048 & 0.055 \\
			\hline
			\end{tabular}
		\end{adjustbox}
	\end{table}
		
		\begin{table}[ht]
		\centering
		\caption{$M2$ assuming NGICA model: Rejection rates in $1000$ repetitions for bootstrap tests of $H_{02}$, $H_{03}$ (true) and $H_{04}$, with $\alpha=0.05$. }
		\label{t6}
		\begin{adjustbox}{width=1\textwidth}
			\begin{tabular}{@{\extracolsep{4pt}}ccccccccccccccccccc@{}}
				\hline
				&\multicolumn{3}{c}{FOBI (boot)}&\multicolumn{3}{c}{\textit{Cov-Cov4}}&\multicolumn{3}{c}{\textit{Cau-Hub}}&\multicolumn{3}{c}{\textit{sCau-sHub}}&\multicolumn{3}{c}{\textit{sCauI-sHubI}} \\
				
				n & $k2$ & $k3$ & $k4$ & $k2$ & $k3$ & $k4$ & $k2$ & $k3$ & $k4$ & $k2$ & $k3$ & $k4$ & $k2$ & $k3$ & $k4$ \\
				\cline{2-4} \cline{5-7} \cline{8-10} \cline{11-13} \cline{14-16} \cline{17-19}
				
				500 & 0.122 & 0.052 & 0.035 & 0.142 & 0.099 & 0.098 & 0.960 & 0.369 & 0.264 & 0.997 & 0.227 & 0.144 & 0.858 & 0.371 & 0.252 \\
  1000  & 0.112 & 0.051 & 0.028 & 0.095 & 0.064 & 0.039 & 0.999 & 0.247 & 0.195 & 1.000 & 0.115 & 0.100 & 0.995 & 0.196 & 0.159 \\
  2000  & 0.086 & 0.021 & 0.014 & 0.064 & 0.029 & 0.021 & 1.000 & 0.121 & 0.107 & 1.000 & 0.050 & 0.036 & 1.000 & 0.086 & 0.114 \\
  4000  & 0.072 & 0.036 & 0.028 & 0.052 & 0.031 & 0.029 & 1.000 & 0.117 & 0.095 &  &  &  & 1.000 & 0.060 & 0.058 \\
				\hline
			\end{tabular}
		\end{adjustbox}
	\end{table}
	
	\begin{table}[ht]
		\centering
		\caption{$M2_x$ assuming NGCA model: Rejection rates in $1000$ repetitions for bootstrap tests of $H_{02}$, $H_{03}$ (true) and $H_{04}$, with $\alpha=0.05.$ }
		\label{t7}
		\begin{adjustbox}{width=1\textwidth}
			\begin{tabular}{@{\extracolsep{4pt}}ccccccccccccccccccc@{}}
				\hline
				&\multicolumn{3}{c}{FOBI (boot)}&\multicolumn{3}{c}{\textit{Cov-Cov4}}&\multicolumn{3}{c}{\textit{Cau-Hub}}&\multicolumn{3}{c}{\textit{sCau-sHub}}&\multicolumn{3}{c}{\textit{sCauI-sHubI}} \\
				
				n & $k2$ & $k3$ & $k4$ & $k2$ & $k3$ & $k4$ & $k2$ & $k3$ & $k4$ & $k2$ & $k3$ & $k4$ & $k2$ & $k3$ & $k4$ \\
				\cline{2-4} \cline{5-7} \cline{8-10} \cline{11-13} \cline{14-16}
				
				500  & 0.264 & 0.069 & 0.033 & 0.403 & 0.200 & 0.095 & 0.923 & 0.110 & 0.083 & 0.993 & 0.263 & 0.105 & 0.763 & 0.237 & 0.130 \\
  1000  & 0.436 & 0.076 & 0.028 & 0.477 & 0.180 & 0.083 & 0.999 & 0.080 & 0.065 & 1.000 & 0.572 & 0.169 & 0.981 & 0.304 & 0.160 \\
  2000  & 0.700 & 0.107 & 0.036 & 0.686 & 0.221 & 0.093 & 1.000 & 0.064 & 0.050 & 1.000 & 0.900 & 0.143 & 1.000 & 0.579 & 0.171 \\
  4000  & 0.934 & 0.066 & 0.009 & 0.919 & 0.071 & 0.047 & 1.000 & 0.060 & 0.038 &  &  &  & 1.000 & 0.871 & 0.131 \\
				\hline
			\end{tabular}
		\end{adjustbox}
	\end{table}

	\begin{table}[ht]
		\centering
		\caption{$M2_x$ assuming NGICA model: Rejection rates in $1000$ repetitions for bootstrap tests of $H_{02}$, $H_{03}$ (true) and $H_{04}$, with $\alpha=0.05$. }
		\label{t8}
		\begin{adjustbox}{width=1\textwidth}
			\begin{tabular}{@{\extracolsep{4pt}}ccccccccccccccccccc@{}}
				\hline
				&\multicolumn{3}{c}{FOBI (boot)}&\multicolumn{3}{c}{\textit{Cov-Cov4}}&\multicolumn{3}{c}{\textit{Cau-Hub}}&\multicolumn{3}{c}{\textit{sCau-sHub}}&\multicolumn{3}{c}{\textit{sCauI-sHubI}} \\
				
				n & $k2$ & $k3$ & $k4$ & $k2$ & $k3$ & $k4$ & $k2$ & $k3$ & $k4$ & $k2$ & $k3$ & $k4$ & $k2$ & $k3$ & $k4$ \\
				\cline{2-4} \cline{5-7} \cline{8-10} \cline{11-13} \cline{14-16}
				
500 & 0.307 & 0.131 & 0.106 & 0.472 & 0.272 & 0.160 & 0.945 & 0.369 & 0.262 & 0.994 & 0.399 & 0.180 & 0.800 & 0.410 & 0.208 \\
  1000 & 0.489 & 0.164 & 0.116 & 0.555 & 0.271 & 0.135 & 0.999 & 0.244 & 0.193 & 1.000 & 0.653 & 0.205 & 0.992 & 0.436 & 0.208 \\
  2000  & 0.757 & 0.200 & 0.129 & 0.736 & 0.286 & 0.136 & 1.000 & 0.157 & 0.107 & 1.000 & 0.929 & 0.143 & 1.000 & 0.671 & 0.157 \\
  4000 & 0.940 & 0.083 & 0.060 & 0.917 & 0.128 & 0.112 & 1.000 & 0.183 & 0.119 &  &  &  & 1.000 & 0.876 & 0.157 \\
				\hline
			\end{tabular}
		\end{adjustbox}
	\end{table}

First we note in Tables \ref{t1}-\ref{t8} that the differences between FOBI and \textit{Cov-Cov4} are rather small and probably mainly due to having different bootstrap samples. At least it is not obvious from these results that the knowledge of the value the eigenvalue of interest is of much relevance. It is however obvious that this combination of scatters does not work well in Model $M2$ as expected due to $s_1$.

Also from the robustness point of view the behaviour is as expected for this scatter combination and it performs poorly in the contaminated settings.
In general it seems that the combination \textit{Cau-Hub} performs best. It works well in uncontaminated and contaminated cases while being more robust than the symmetrized counterparts. This is not a surprise as outliers have larger effects when symmetrizing and especially in the incomplete case. Comparing the symmetrized and incomplete symmetrized results it can be seen that the incomplete case starts to work in the uncontaminated settings when the sample sizes are sufficiently large, which is acceptable as it would anyway only be used when the usage of all pairwise differences would become too costly.

The knowledge whether the data actually follows an NGCA model or an NGICA model during bootstrap seems also only of minor relevance while the results in the NGICA model seem to be slightly worse then in the broader NGCA model, which can be simply due to difference in bootstrap samples. However, it is also possible that the difference in performance of bootstrap tests wrongly assuming NGICA and assuming NGCA would be larger in data sets where more dependence is introduced into signal components.\\
	
In Section \ref{Radojicic:sec:4} we suggested a strategy for testing the dimension of the signal space in NGCA and NGICA using any pair of scatter matrices. The simulation results show that under the null hypothesis of exactly $k=q$ non-Gaussian components, the alpha level is kept while the rejection frequencies are low if $k$ is larger than $q$ and high if $k$ is smaller than $q$. This is in accordance with Result~\ref{th1} which was derived however for FOBI only.

\section{Estimation of the signal space dimension}
\label{Radojicic:sec:6}

Usually the dimension $q$ in NGCA or NGICA is unknown and therefore needs to be estimated from the data. The results from Section \ref{Radojicic:sec:5} encourage us to apply for this purpose the hypothesis tests successively. Different strategies for the successive testing are possible and while the test statistic is monotone in the dimension, its distribution is changing as can be seen from the FOBI results. Therefore different strategies might not yield the same dimension estimate.\\

In the following we will introduce two different strategies and compare them in a simulation study. The first strategy is denoted as the incremental strategy. This strategy assumes initially at least one Gaussian component and then tests successively, at level $\alpha$, $H_{0k},\,\,k=p-2,\ldots,0$. The estimated $\hat q$ is the smallest $k$ for which $H_{0k}$ is not being rejected at level $\alpha$, i.e.

$$
\hat{q}=\min\{ k\in\{0,\ldots,p-2\}:H_{0k} \text{ is not being rejected}\}.
$$

An algorithmic scheme is presented for this strategy in  Algorithm~\ref{alg::est}.

	\begin{algorithm}[htb]
	Set the proposed dimension $k=p-2$;\;\\
	Set the significance level $\alpha$;\;\\
	Initiate the parameters of the Algorithm~\ref{alg::boot};\;\\
	\Repeat{$\hat{p}_k\leq\alpha$ or $k=0$}{
		Test for $H_{0k}$ and compute bootstrap p-value $\hat{p}_k$ using Algorithm~\ref{alg::boot};\;\\
		\If{$\hat{p}_k>\alpha$}{$k=k-1$\;}}

	Return the estimate $\hat{q}=k+1$ of the signal dimension\;
	\caption{Estimating dimension $q$ of the signal subspace using an incremental approach}\label{alg::est}
\end{algorithm}

For the incremental strategy the number of Gaussian components should be preferably small. If one suspects that this would not be the case, for example a divide and conquer strategy could be applied to find a point where acceptance switches to rejection at a specific level $\alpha$. A possible variant for a divide and conquer strategy is presented in Algorithm~\ref{alg::dic}.
\begin{algorithm}[htb]
	Set the proposed dimension $k=\lceil \frac{p}{2}\rceil$;\;\\
	Set the significance level $\alpha$;\;\\
	Set $q_{min}^0=1$ and $q_{max}^0=p-1$;\;\\
	Initiate the parameters of the Algorithm \ref{alg::boot};\;\\
	
	\Repeat{$q_{min}^0=q_{max}^0$}
	{
		Test $H_{0k}$ and $H_{0(k-1)}$ using Algorithm \ref{alg::boot};\;\\
		\If {$H_{0k}$ is not rejected and $H_{0(k-1)}$ is rejected }{ Return $\hat{q}=k$;\;\\}
		\If {$H_{0k}$ is not rejected and $H_{0(k-1)}$ is not rejected }{ $q_{min}^1\leftarrow q_{min}^0$, $q_{max}^1\leftarrow k-1$, $k=\lceil \frac{q_{max}^1+q_{min}^1}{2}\rceil$\;\\}
		\If {$H_{0k}$ is rejected}{$q_{min}^1\leftarrow k+1$, $q_{max}^1\leftarrow q_{max}^0$, $k=\lceil \frac{q_{max}^1+q_{min}^1}{2}\rceil$\;\\}
		Update:\,\, $q_{min}^0\leftarrow q_{min}^1$, $q_{max}^0\leftarrow q_{max}^1$\;\\
	}
	Return the estimate $\hat{q}=k$ of the signal dimension\;
	\caption{Estimating dimension $q$ of the signal subspace using divide and conquer strategy}\label{alg::dic}
\end{algorithm}

Naturally in both algorithms prior knowledge could be incorporated by adjusting the starting points of the procedures and also many other strategies are possible. As suggested in \cite{Nordhausen:2016}, a sequence of bootstrap test sizes $\alpha_k$ for testing $H_{0k}$ can be determined so that the consistency of the procedure is preserved, but due to simplicity we will use fixed test sizes $\alpha_k=\alpha=0.05, \,\forall k$.\\

We restrict ourselves to compare only these two strategies by adjusting models $M1$ and $M2$ slightly. In the adjusted models $M1^*$ and $M2^*$ the same signal components are used as in $M1$ and $M2$ respectively, but the number of Gaussian components is doubled to $6$. As there was little difference in performance when bootstrapping an underlying NGCA or NGICA model, 
we restrict ourselves to assume an NGCA model. Moreover, encouraged by results presented in Tables \ref{t1}-\ref{t8} we compare only the scatter combinations \textit{Cov-Cov4} and \textit{Cau-Hub}, where all tests are executed at level $\alpha=0.05$.\\

\begin{figure}[ht]
 	\centering
 	\includegraphics[width=1\textwidth,keepaspectratio]{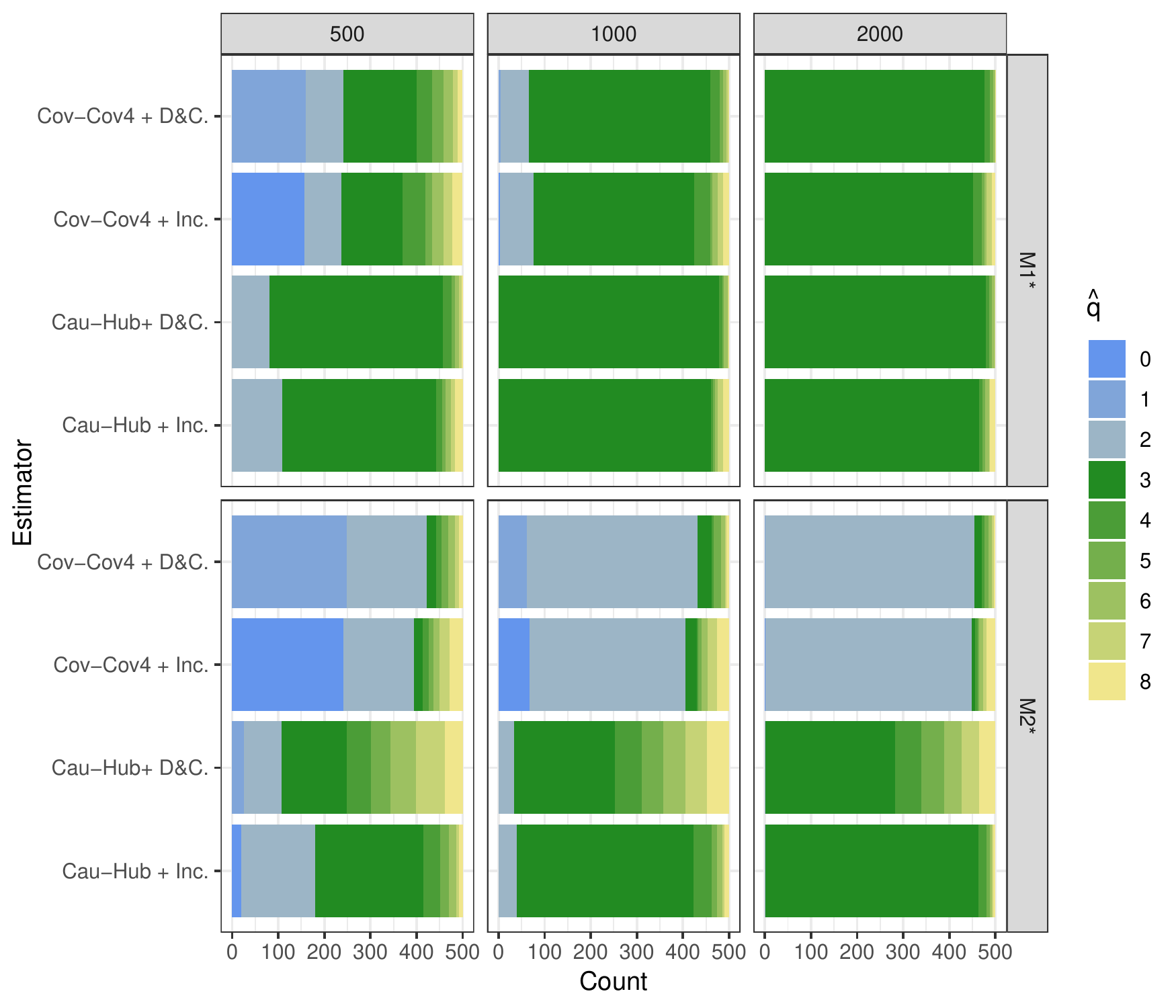}
 	\caption{Frequencies of estimated dimension of the non-Gaussian subspace for incremental strategy (Inc) and divide and conquer (D\&C) strategy in models $M1$* and M2* based on 500 iterations when using different scatter combinations and different sample sizes.}
 	\label{fig:est_q}
 \end{figure}

Based on $500$ repetitions Figure \ref{fig:est_q} shows the estimated signal dimensions.

The Figure  \ref{fig:est_q} shows that especially with increasing sample size correct dimensions are estimated in both models when using \textit{Cau-Hub}, whereas as expected, \textit{Cov-Cov4} fails to recognize one signal in $M2^*$. It needs however also larger sample sizes compared to \textit{Cau-Hub} in model $M1^*$. It also shows that there are differences between the strategies and at least here incremental strategy looks a bit better, which could possibly be justified by argumentation presented before in this section.

\section{Conclusion}
	
Dimension reduction is of increasing importance and quite often it is
considered that the interesting subspace of the data is non-Gaussian.
NGCA and NGICA are two dimension reduction approaches which follow these
ideas and try to separate the Gaussian subspace from the non-Gaussian
one. There are many methods suggested in the literature for NGCA
and NGICA but usually they assume that the dimensions of the subspaces
are known, which is rather unrealistic. In this paper we show under which
conditions two different scatter matrices can be used to estimate the
subspaces. Based on this approach we suggest also bootstrap tests to
test for a specific subspace dimension and show how successive
applications of the presented tests can be used to obtain an estimate of the
dimensions of interest. A disadvantage of our suggestion is the
computational complexity which also depends on the scatter matrices
selected. Especially when using symmetrized scatters this becomes quite
demanding, but if the sample sizes are large it seems that incomplete
symmetrized scatters can be successfully used too. However, as we pointed out -
usage of symmetrized scatters is actually not required if the goal is just to separate the two
subspaces, since also non-symmetrized scatters can be rightfully used for the
separation. It is just in the NGICA model that these combinations might not be able
to recover the signals. Therefore, one strategy here could be to use
computationally faster and often more robust regular scatter functionals in order to find the
non-Gaussian subspace, and then to apply, on the estimated subspace, a regular ICA
method, for example one based on two symmetrized scatter matrices, to
estimate the independent components.

\section{Appendix}

\textbf{Proof of the Result \ref{res1}} Assume $\bo x$ follows an NGCA model formulated using location functional $\bo T$ and scatter functional $\bo S_1$, $\bo x=\bo A\bo z=\bo A_1\bo s+\bo A_2\bo n$, and let $\bo S_2$ be  scatter functional different from $\bo S_1$. 

\noindent Let $\bo S_2(\bo x^{st})=\tilde{\bo U}\bo D\bo {\tilde{\bo U}}^\top$ be eigen decomposition of $\bo S_2$, where $\bo x^{st}=\bo S_1(\bo x)^{-1/2}\bo x$ and the eigenvalues in $\bo D$ are ordered so that $d_1>\cdots>d_q$ and $d_{q+1}=\cdots=d_p$. Let $\bo W=\tilde{\bo U}^\top\bo S_1(\bo x)^{-1/2}$ and $\bo A=\bo U\bo L\bo V$ be an SVD decomposition of mixing matrix $\bo A$. Since $\bo x=\bo A\bo z$,
$$
{\bo S_1(\bo x)}^{-1/2}\bo x=\bo U\bo V^\top\bo z,\quad \bo S_2(\bo x^{st})=\bo U\bo V^\top\bo S_2(\bo z)(\bo U\bo V^\top)^\top.
$$
$\bo S_2(\bo z)$ and $\bo S_2(\bo x^{st})$ are similar and thus have the same eigenvalues. Hence
$$
\bo S_2(\bo z)=\bo U_B\bo D\bo U_B^\top \implies \bo S_2(\bo x^{st})=\bo U\bo V^\top\bo U_B\bo D\bo U_B^\top(\bo U\bo V^\top)^\top,
$$
where $\bo U_B$ is orthogonal matrix. Since $\bo S_2(\bo x^{st})=\tilde{\bo U}\bo D\bo {\tilde{\bo U}}^\top$ then $\tilde{\bo U}=\bo U\bo V^\top\bo U_B\bo P_B\bo J$, where $\bo J$ is a sign-changing matrix and $\bo P_B=\text{diag}(\bo I_q,\bo P_{p-q})$ is block-diagonal matrix with the first block being an identity and the second block being a permutation matrix. Therefore
$$
\bo W=(\bo U\bo V^\top\bo U_B\bo P_B\bo J)^\top {\bo S_1(\bo x)}^{-1/2}.
$$
$\bo S_2(\bo z)$ is a block-diagonal matrix implying that $\bo U_B$ is also block-diagonal, with orthogonal blocks $\bo U_{B1}\in\mathbb{R}^{q\times q}$ and $\bo U_{B2}\in\mathbb{R}^{(p-q)\times (p-q)}$. Hence,
$$
\bo W\bo x=\bo J^\top(({\bo U_{B1}^\top\bo s})^\top\,({\bo P_{p-q}^\top\bo U_{B2}^\top\bo n})^\top)^\top.
$$
\\

\noindent\textbf{Proof of the Result \ref{res2}} Assume $\bo x$ follows an NGICA model formulated using location functional $\bo T$ and scatter functional $\bo S_1$ with block-independence property, $\bo x=\bo A\bo z=\bo A_1\bo s+\bo A_2\bo n$, and let $\bo S_2$ be scatter functional different from $\bo S_1$ also having block-independence property. 

\noindent Let $\bo S_2(\bo x^{st})=\tilde{\bo U}\bo D\bo {\tilde{\bo U}}^\top$ be eigen-decomposition of $\bo S_2(\bo x^{st})$, where $\bo x^{st}=\bo S_1(\bo x)^{-1/2}\bo x$ and the eigenvalues in $\bo D$ are ordered so that $d_1>\cdots>d_q$ and $d_{q+1}=\cdots=d_p$. Let $\bo W=\tilde{\bo U}^\top\bo S_1(\bo x)^{-1/2}$ and $\bo A=\bo U\bo L\bo V$ be an SVD decomposition of mixing matrix $\bo A$. Since $\bo x=\bo A\bo z$,
$$
{\bo S_1(\bo x)}^{-1/2}\bo x=\bo U\bo V^\top\bo z,\quad \bo S_2(\bo x^{st})=\bo U\bo V^\top\bo S_2(\bo z)(\bo U\bo V^\top)^\top.
$$
$\bo S_2(\bo z)$ and $\bo S_2(\bo x^{st})$ are similar and thus have the same eigenvalues. Hence
$$
\bo S_2(\bo z)=\bo U_B\bo D\bo U_B^\top \implies \bo S_2(\bo x^{st})=\bo U\bo V^\top\bo U_B\bo D\bo U_B^\top(\bo U\bo V^\top)^\top,
$$
where $\bo U_B$ is orthogonal matrix. Since $\bo S_2(x^{st})=\tilde{\bo U}\bo D\bo {\tilde{\bo U}}^\top$ then $\tilde{\bo U}=\bo U\bo V^\top\bo U_B\bo P_B\bo J$, where $\bo J$ is a sign-changing matrix and $\bo P_B=\text{diag}(\bo I_q,\bo P_{p-q})$ is block-diagonal matrix with the first block being an identity and the second block being a permutation matrix. Therefore
$$
\bo W=(\bo U\bo V^\top\bo U_B\bo P_B\bo J)^\top {\bo S_1(\bo x)}^{-1/2}.
$$
$\bo S_2(\bo z)$ is a block-diagonal matrix implying that $\bo U_B$ is also block-diagonal, with orthogonal blocks $\bo I_q \in\mathbb{R}^{q\times q}$ and $\bo U_{B2}\in\mathbb{R}^{(p-q)\times (p-q)}$. Hence,
$$
\bo W\bo x=\bo J^\top({\bo s}^\top\,({\bo P_{p-q}^\top\bo U_{B2}^\top\bo n})^\top)^\top.
$$
\\

\noindent\textbf{Proof of the Result \ref{res3}}
 Assume $\bo x$ follows an NGICA model, $\bo x=\bo A\bo z=\bo A_1\bo s+\bo A_2\bo n$, and assume that all but one of one component of $\bo s$ are symmetric. Since $\bo n$ has Gaussian distribution, all but one of the independent blocks in $\bo z$ are symmetric implying that any scatter matrix $\bo S(\bo z)$, provided that it exists at $\bo z$, has the block-independence property. Now, the Result \ref{res3} follows directly from Result \ref{res2}.

\section*{Acknowledgement}
The work of KN was supported by the Austrian Science Fund (FWF) Grant number P31881-N32.

\end{document}